\documentclass[twocolumn,showpacs,preprintnumbers,amsmath,amssymb]{revtex4}

\usepackage{graphicx}
\usepackage{dcolumn}
\usepackage{bm}

\begin{document}

\preprint{APS/123-QED}

\title{The intruder feature of $^{31}{\bf Mg}$ and \\the
coexistence of many particle and many hole states }
\author{M. Kimura}
\affiliation{Institute of Physics, University of Tsukuba, Tsukuba 305-8571, Japan<}
\date{\today}

\begin{abstract}
The low-lying level structure of $^{31}{\rm Mg}$ has been investigated
by the antisymmetrized molecular dynamics (AMD) plus
generator coordinate method (GCM) with the Gogny D1S force. It is
shown that the N=20  magic number is broken and the ground state
has the pure neutron $2p3h$ configuration. The coexistence of many
particle and many hole states at very low excitation energy is
discussed.
\end{abstract}

\pacs{Valid PACS appear here}
\maketitle

The breakdown of the neutron magic number  N=20 and the  change of
the shell order in neutron rich nuclei have been one of the great interest
in nuclear structure studies. Experimentally, it was firstly pointed
out from the observation of the anomalous binding energy and spin-parity
of $^{31}{\rm Na}$ \cite{Thibault,Huber}. More convincing  
evidence was given by the observation of the small excitation energy 
\cite{Detraz,Guillemaud} and the large $E2$ transition probability
\cite{Motobayashi} of the $2^+_1$ state of $^{32}{\rm
Mg}$. Theoretically, the Hartree-Fock calculation \cite{Campi} showed that shape
transition from spherical to prolate shapes possibly took place in
neutron-rich N=20 nuclei.  Shell model calculations \cite{Poves,
Warburton, Fukunishi} that allowed particle hole excitations across the
N=20 shell gap explained these abnormal properties and suggested 
strong deformation of  $^{31}{\rm Na}$ and neighboring nuclei caused by
the inversion between the normal and intruder configurations.  Since
then, many experimental \cite{Klotz, Pritychenko, Keim, Yoneda, Nummela,
Yanagisawa, Neyens, Marechal} and theoretical \cite{Wildenthal, Poves2,
Utsuno1, Utsuno2, Dean, Rodriguez, Caurier1, Caurier2, Kimura2, Kimura3,Yamagami}
studies have been made and now the systematic breaking of N=20 magic
number is theoretically investigated and some of them are
observed. Thus, our knowledge of the breakdown of the neutron magic
number N=20 in the neutron rich nuclei has been increasing rapidly in
this decade.   

 To understand the shell structure and the shell order, the odd-neutron
nuclei such as $^{31}{\rm Mg}$ and $^{29}{\rm Ne}$ will provide
essential information, since the property of the single particle orbital
of the last neutron particle or hole will be strongly reflected to the
property of total system. Recently the spin-parity of the ground state
of $^{31}{\rm Mg}$ was experimentally fixed 
to $1/2^+$ \cite{Neyens}. This result suggests that the 
ground state of $^{31}{\rm Mg}$ has the intruder configuration in
which two neutron occupy the $pf$-orbital and a neutron hole
is in the Nilsson orbital [200,1/2].  However, mainly because
of the limitation of the models,  theoretical investigation of
odd-neutron nuclei is not as vigorous as that of even-even
nuclei. Indeed, there are only a few shell model calculations
\cite{Wildenthal,Poves2,Neyens,Marechal} for $^{31}{\rm Mg}$, and they
do not reproduce the ground state properties or they study the ground
band and a few excited states using the modified interaction to
reproduce the ground state properties. 

In this work, we discuss the structure of the $^{31}{\rm Mg}$ based on
the theoretical framework of AMD+GCM \cite{Enyo1,Kimura1}. Since we do not assume
the time reversal symmetry and we perform the parity projection before
the variation, we can appropriately investigate the positive- and negative-parity
states and the coexistence of the many particle-hole states in
odd-mass nuclei. It will be shown that the ground state of $^{31}{\rm
Mg}$ has the pure intruder configuration that has the $2p3h$
configuration with respect to the N=20 shell closure. We have
successfully reproduced the observed properties of $^{31}{\rm Mg}$ such
as the magnetic moment and the $\beta$-decay strengths. We also predict
that other many particle-hole states ($0p1h$, $1p2h$ and
$3p4h$) also coexist at very small excitation energy. 

In the AMD, the intrinsic wave function  is given by a Slater
determinant of the single particle wave packets,
\begin{eqnarray*}
 \Phi_{\rm int} &=& {\mathcal A}
  \{\varphi_1,\varphi_2,...,\varphi_A \} ,\quad
 \varphi_i({\bf r}) = \phi_i({\bf r})\chi_i\xi_i ,
\end{eqnarray*}
where $\varphi_i$ is the $i$th single particle wave packet
consisting of the spatial $\phi_i$, spin $\chi_i$ and isospin
$\xi_i$ parts. The local Gaussian located at ${\bf Z}_i$ is employed as
$\phi_i$.  
\begin{eqnarray*}
 \phi_i({\bf r}) &=& \exp\biggl\{-\sum_{\sigma=x,y,z}\nu_\sigma
  \Bigl(r_\sigma - 
  \frac{{\rm Z}_{i\sigma}}{\sqrt{\nu_\sigma}}\Bigr)^2\biggr\}, 
  \nonumber\\ 
 \chi_i &=& \alpha_i\chi_\uparrow + \beta_i\chi_\downarrow,
  \quad |\alpha_i|^2 + |\beta_i|^2 = 1,\nonumber \\
 \xi_i &=& {\rm proton} \quad {\rm or} \quad {\rm neutron}. 
\end{eqnarray*}
Here the centroids of the Gaussian ${\bf Z}_i$, the spin direction
$\alpha_i$ and $\beta_i$ and the width parameters $\nu_x, \nu_y$
and $\nu_z$ are the variational parameters. As the variational wave
function, we employ the parity projected wave function,
$ \Phi^{\pi} = \hat{P}^\pi \Phi_{\rm int}$. The Gogny D1S force
\cite{Gogny} is employed as an effective nuclear force
$\hat{V}_n$. Coulomb force  $\hat{V}_c$ is approximated by the sum of
seven Gaussians.  

The calculation is performed in the following three steps. First, we
carry out the energy variation and optimize the variational
parameters. The energy variation is made under the constraint on the
matter quadrupole deformation $\beta$ to obtain the optimum wave
function $\Phi^\pi(\beta)$ for given value $\beta$. We note that in
this calculation, we do not make any assumption on the spatial symmetry
of the wave function and do not put any constraint on the quadrupole
deformation parameter $\gamma$. Therefore, $\gamma$ has the optimal
value for each given value of $\beta$. The definition of $\beta$ and
$\gamma$ is given in Ref. \cite{Dote}. Then, we project out the
eigenstate of the total angular momentum $\hat{J}$ from the optimized
wave function, $ \Phi^{J\pi}_{MK}(\beta) =
\hat{P}^{J}_{MK}\Phi^{\pi}(\beta)$.  
$\Phi^{J\pi}_{MK}(\beta)$ is referred as AMD wave function in the
following. Finally, we superpose all of the AMD wave functions on the
energy surface that have the same parity and the angular momentum but
have different $K$ and $\beta$, 
\begin{eqnarray*}
 \Phi_n^{J\pi} = c_n\Phi^{J\pi}_{MK}(\{\beta\})
  + c_n^\prime\Phi^{J\pi}_{MK^\prime}(\{\beta^\prime\}) + \cdots.
\end{eqnarray*}
 The coefficients $c_n$, $c'_n$,... are
determined by solving the Hill-Wheeler equation
\cite{Hill}. $\Phi_n^{J\pi}$ will be referred as AMD+GCM wave function.

We have investigated the single particle configuration of each intrinsic
wave function by the AMD+HF method \cite{Dote}. Once we obtain the
intrinsic wave function, 
$\Phi_{\rm int}(\beta)$,  we transform the single particle wave packet
$\varphi_i$ into the orthnormalized basis $\widetilde{\varphi}_\alpha$, 
\begin{eqnarray*}
 \widetilde{\varphi}_\alpha = \sum_{i=1}^A
  \frac{1}{\sqrt{\lambda}_\alpha}c_{i\alpha}\varphi_i,
\end{eqnarray*}
where $\lambda_\alpha$ and $c_{i\alpha}$ are the eigenvalue and
eigenvector of the matrix $B_{ij}$=$\langle\varphi_i|\varphi_j\rangle$.
From this new basis function, we construct and diagonalize the
Hartree-Fock single particle Hamiltonian,
\begin{eqnarray*}
 h_{\alpha\beta} &\equiv&
  \langle\widetilde{\varphi}_\alpha|\hat{t}|\widetilde{\varphi}_b\rangle
  +  \sum_{\gamma=1}^{A}\langle
  \widetilde{\varphi}_\alpha\widetilde{\varphi}_\gamma
  |{\hat{v}_n+\hat{v}_c}|\widetilde{\varphi}_\beta 
  \widetilde{\varphi}_\gamma -
  \widetilde{\varphi}_\gamma\widetilde{\varphi}_\beta\rangle,\nonumber\\
 &+&\frac{1}{2}\sum_{\gamma,\delta=1}^{A}\langle
  \widetilde{\varphi}_\gamma\widetilde{\varphi}_\delta 
  |\widetilde{\varphi}_\alpha^*\widetilde{\varphi}_\beta
  \frac{\partial\hat{v}_n}{\partial \rho}|\widetilde{\varphi}_\gamma
  \widetilde{\varphi}_\delta - \widetilde{\varphi}_\delta
  \widetilde{\varphi}_\gamma \rangle,\\
 h_{\alpha\beta} f_{\beta s} &=& \epsilon_sf_{\alpha s}.
\end{eqnarray*}
The eigenvalue and eigenvector provide the single particle energy
$\epsilon_s$ and the single particle wave function $\psi_s$ =$\sum_{\alpha=1}^A
f_{\alpha s}\widetilde{\varphi}_\alpha$. To investigate the properties
of the single particle wave functions, we also calculate the magnetic
quantum number $\langle\psi_s|\hat{j_z}|\psi_s\rangle$ and the amount of
the positive-parity component
$\langle\psi_s|\hat{P}^+|\psi_s\rangle$. We note that by this method we 
obtain only the occupied states. In the following, we denote the neutron
$m$ particle and $n$ hole configuration relative to the N=20 spherical
shell closure as $mpnh$.  
\begin{figure}[bt]
 \includegraphics[width=\hsize]{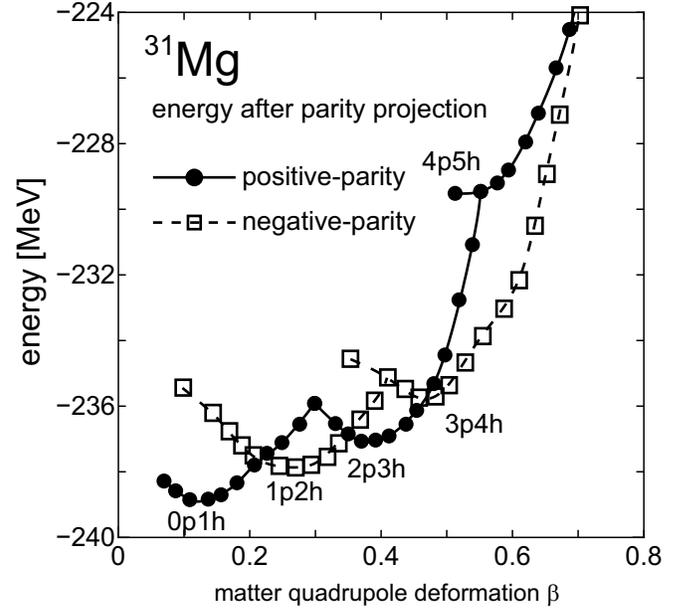}
\caption{Energy surfaces of $^{31}{\rm Mg}$ as functions of matter
 quadrupole deformation parameter $\beta$ for the positive- and
 negative-parity states. The estimated neutron particle hole
 configurations are  also shown around the energy
 minima.}\label{fig::surface31Mg}  
\end{figure}

\begin{figure}
 \includegraphics[width=\hsize]{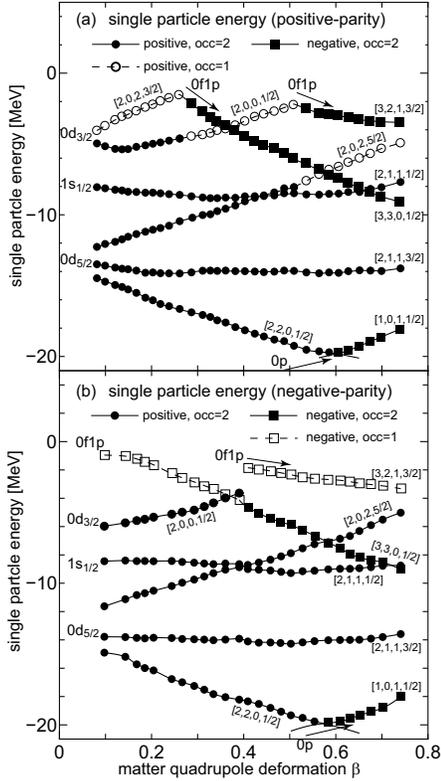}
\caption{The single particle orbitals of the last 11 neutrons of the
 positive-parity state (a) and the negative-parity state (b). Filled
 (open) symbols show the orbitals occupied by two (one)
 neutrons. Circles (boxes) show the orbital in which the amount of the
 positive-parity component is larger than (smaller than) 50\%.}
\label{fig::spo31Mg}
\end{figure}
Fig. \ref{fig::surface31Mg} shows the calculated energy surface of
the positive- and negative-parity states. There are three local minima in
the positive-parity state and two local minima in the negative-parity
state. Though we do not assume the axial symmetry, the wave functions on
the energy surfaces are almost axially symmetric except for those located
at the transitional region between the energy minima. These minima have
different neutron single particle
configurations. Fig. \ref{fig::spo31Mg} shows the single particle
orbitals occupied by the last 11 neutrons in the positive-parity state 
(a) and the negative-parity state (b). In this figure, the asymptotic
quantum number $[N,n_z,l_z,j_z]$ of each orbital evaluated from its
parity and magnetic quantum number is also shown. In the following,
we use the asymptotic quantum number to specify each neutron
orbital. Fig. \ref{fig::spo31Mg} (a) shows that in the spherical region
of the positive-parity state, the last neutron particle or the neutron
hole occupies [2,0,2,3/2] orbital that originates in the
$0d_{3/2}$-orbital. Therefore this configuration is understood as the $0p1h$ 
configuration. Around $\beta\sim0.3$, the neutron configuration changes
from $0p1h$ to $2p3h$. The last two neutron occupy [3,3,0,1/2]
orbital that intrudes from the $pf$-orbital. And a neutron hole occupies
[2,0,0,1/2] orbital that originates in the $0d_{3/2}$-orbital. Around
$\beta\sim 0.5$, the neutron configuration changes from $2p3h$ to
$4p5h$. [2,0,2,3/2] and [2,0,0,1/2] orbitals are no longer
occupied. Both of [3,3,0,1/2] and [3,2,1,3/2] orbitals intruding from
the $pf$-orbital are occupied by two neutrons. A neutron hole is in
[2,0,2,5/2] orbital that originates in the $0d_{5/2}$-orbital. In the
same way, Fig. \ref{fig::spo31Mg} (b) shows that there are two different
neutron configurations $1p2h$ and $3p4h$ in the negative-parity state.
The $1p2h$ configuration has the last neutron in [3,3,0,1/2] orbital and
[2,0,2,3/2] orbital is unoccupied. Around $\beta\sim 0.4$ the
configuration changes from $1p2h$ to $3p4h$. The $3p4h$ configuration
has the last neutron in [3,2,1,3/2] orbital and both of [2,0,2,3/2] and
[2,0,0,1/2] orbitals are unoccupied. Thus,  the analysis of the single
particle configurations has revealed that the minima of the
positive-energy state correspond to the neutron $0p1h$, $2p3h$ and
$4p5h$ configurations and those of the negative-parity state correspond
to the neutron $1p2h$ and $3p4h$ configurations. It is also mentioned
that  protons are always below the Z=20 shell closure and do not excited
from the $0p$-orbital to the $1s0d$-orbital, though the order of the
proton [2,2,0,1/2] and [1,0,1,1/2] orbitals has changed in the strongly
deformed region. Therefore, the parity of total system depends only
on the neutron configuration. There are some additional comments on
Fig. \ref{fig::surface31Mg} and \ref{fig::spo31Mg}. (1) All of the
$mpnh$ configurations except for the $4p5h$ configuration appear at very
small excitation energy.  The energy difference between the lowest one
($0p1h$) and the highest one ($3p4h$) is less than 5 MeV. This leads to
the coexistence of these particle-hole configurations within small
excitation energy after the angular momentum projection as shown
later. (2) The behavior of the neutron single particle orbitals of the
positive- and negative-parity states are qualitatively similar to each
other. As deformation becomes larger, [3,3,0,1/2] and [3,2,1,3/2]
orbital originate in the $pf$-orbital come down and [2,0,2,3/2] and
[2,0,0,1/2] orbital originate in the $0d_{3/2}$-orbital go up and cross 
to each other. The change of the particle hole configuration occurs just
around the crossing points of these orbitals. This explains why the
particle hole configurations appear in the order of the particle and
hole numbers as deformation becomes larger. It is also mentioned that
the behavior of the single particle orbital is qualitatively understood
by the Nilsson model and it may explain why the Nilsson model
calculation was qualitatively successful in Ref. \cite{Marechal}. (3) The
energy of the $4p5h$ configuration is exceptionally higher than others
and it is understood as follows. As explained above, the change of the
neutron particle-hole configuration is caused by the crossing of four
neutron orbitals that originate in the $pf$- and
$0s1d$-orbitals. Therefore, up to four particle or four hole
configuration can appear at small excitation energy, but more than five
particle or hole configuration  appear at rather high excitation
energy. This feature is common to other neighboring nuclei such as Ne
isotopes \cite{Kimura2, Enyo2} and Mg isotopes \cite{Kimura3}.    

\begin{figure}
 \includegraphics[width=\hsize]{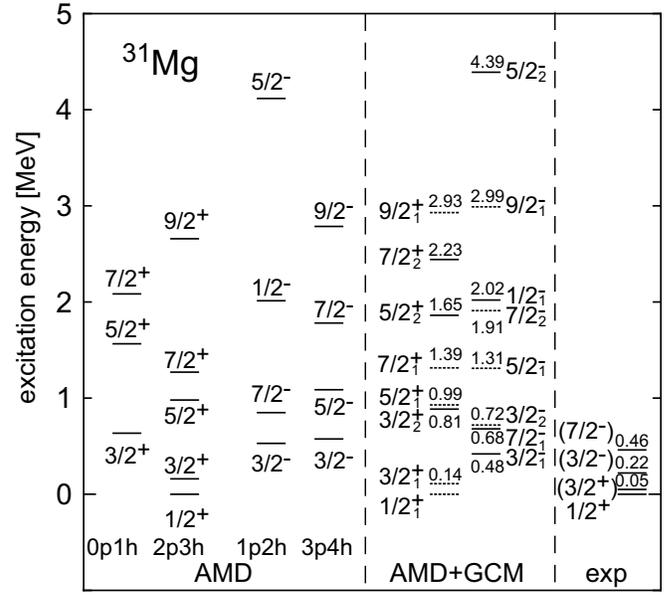}
\caption{Level scheme of $^{31}{\rm Mg}$ obtained by the AMD wave
 functions at energy minima and by the AMD+GCM calculation together with
 the partial level scheme suggested in Ref. \cite{Neyens}. 
 Solid (dashed) lines show the states dominated by the neutron $0p1h$
 or $1p2h$ ($2p3h$ or $3p4h$) configurations. Numbers above or blow the
 states show the  calculated and observed excitation energy in MeV.}
 \label{fig::level31Mg} 
\end{figure}
\begin{table*}[tbh]
\begin{ruledtabular}
\begin{tabular}{ccccccccc}
 \multicolumn{5}{c} {decay from $^{31}{\rm Na}(\rm g.s.)$} &
 \multicolumn{4}{c} {decay to $^{31}{\rm Al}$} \\
$^{31}{\rm Mg}(J^\pi)$ & $E_x$ (MeV) & $\log ft$ & $0p1h$ (\%) & $2p3h$ (\%) &
$^{31}{\rm Al}(J^\pi)$ & $E_x$ (MeV) & $\log ft$ & $2p4h$ (\%)\\
\cline{1-5}\cline{6-9}
$1/2^+_1$ & 0    & 4.6 & 0 & 93  & $1/2^+_1$ & 1.22 & $6.1$ & 5 \\
$3/2^+_1$ & 0.14 & 5.2 & 25 & 61 & $3/2^+_1$ & 1.81 & 5.8 & 10 \\
$3/2^+_2$ & 0.81 & 5.8 & 68 & 20 & $1/2^+_2$ & 3.08 & 5.2 & 69 \\
$5/2^+_1$ & 0.89 & 5.3 & 21 & 63 & $3/2^+_2$ & 3.83 & 5.0 & 61 \\
$5/2^+_2$ & 1.85 & 5.9 & 67 & 11 & $3/2^+_3$ & 4.11 & 4.8 & 79 \\
          &      &     &    &    & $1/2^+_3$ & 4.43 & 5.2 & 67 \\
\end{tabular}
\caption{The calculated $\log ft$ values of the $\beta$-decay from the
 ground state  of $^{31}{\rm Na}$ to the low-lying states of $^{31}{\rm
 Mg}$ and from  the ground state of $^{31}{\rm Mg}$ to the low-lying
 states of  $^{31}{\rm Al}$. } \label{tab::logft}
\end{ruledtabular}
\end{table*}
After the variational calculation, we have performed the angular
momentum projection of the wave functions on the energy surfaces.
The left side of FIG. \ref{fig::level31Mg} shows the excitation energies
of the AMD wave functions at the energy minima. In the case of the
positive-parity state, the $0p1h$ minimum generates the $3/2^+$, $5/2^+$
and $7/2^+$ states and the $2p3h$ minimum generates the $K^\pi$=$1/2^+$
rotational band with negative decoupling parameter. The $4p5h$ minimum
generates the $K^\pi$=$5/2^+$  rotational band, but it is not shown
since it is not involved in the low-lying states even after the GCM
calculation. In the case of the negative-parity state, $1p2h$ minimum
generates the $K^\pi$=$1/2^-$ rotational band with the large negative
decoupling parameter $a\sim-4$ (the last neutron that occupy the
$0f_{7/2}$ gives $a=-4$). It is interesting that the lowest state of
$1p2h$ configuration is not $7/2^-$ but $3/2^-$ because of its
deformation.  The $3p4h$ configuration generates the $K^\pi$=$3/2^-$
rotational band. As clearly seen, the energy of the deformed states
becomes lower. For example, the $3/2^-$ state that has the $3p4h$
configuration is lowered by about 5.1 MeV from the minimum of the energy
surface, while the $3/2^+$ state that has the $0p1h$ configuration is
lowered only by about 1.8 MeV. This leads to the coexistence of many
particle-hole states within quite small excitation energy. However, it
should be noted that  there is no $1/2^+$ state except for the $2p3h$
configuration. It is also mentioned that the $K$ quantum number of the
deformed bands ($K^\pi$=$1/2^+$, $1/2^-$ and $3/2^-$) correspond to the
asymptotic quantum number $j_z$ of the last neutron particle or hole.

To complete our calculation, we have performed the GCM calculation by
superposing all of the AMD wave functions on the energy surfaces. The
calculated level scheme is shown in the middle of
FIG. \ref{fig::level31Mg}. There are mixing between different
particle-hole configurations. For example, in the case of the $3/2^+$
states, the mixing between the $0p1h$ and $2p3h$ configurations lowers
the energy of the $3/2^+_1$ state that has the dominant $2p3h$
configuration, while it pushes up the $3/2^+_2$ state that has the
dominant $0p1h$ configuration. Since the AMD+GCM wave function is the
sum of many Slater determinants, its particle-hole configuration is
ambiguous. To evaluate it, we have calculated the squared overlap
between the AMD+GCM wave function of each state and the AMD wave
function at each energy minima that has the definite particle-hole
configuration.  The energy gain by such configuration mixing does not
exist in the case of the $1/2^+_1$ state that purely consist of $2p3h$
configuration (it has no overlap with the $0p1h$ AMD wave function but
has 93\% overlap with the $2p3h$ AMD wave function), but it is the ground state 
with a small margin and  consistent with the experiment
\cite{Neyens}. The calculated total binding energy is 242.9 MeV.
The excitation spectrum also shows the good
agreement with the most recent spin-parity assignment \cite{Neyens}. The
first excited state is the $3/2^+_1$ state 
that has the dominant $2p3h$ configuration. The squared overlap with the
$2p3h$ AMD wave function amounts to 61\%. The low-lying
negative-parity states appear as the second ($3/2^-_1$), third
($7/2^-_1$) and fourth ($3/2^-_2$) excited states with the dominant $1p2h$, 
$1p2h$ and $3p4h$ configurations, respectively. The normal configuration
($0p1h$) appears as the fifth excited state $3/2^+_2$ at 0.81 MeV. Thus,
from $0p1h$ to $3p4h$ states, many particle-hole configurations coexist
below 1 MeV. 
\begin{table}[tbh]
\begin{ruledtabular}
\begin{tabular}{cccccc}
$J^\pi$ & $Q$ & $Q_s$ & $J^\pi$ & $Q$ & $Q_s$ \\\cline{1-3}\cline{4-6}
$3/2^+_1$ & -18.9  & -17.1 & $3/2^+_2$ & 8.2 & 7.4  \\
$5/2^+_1$ & -19.1 & -21.6 & $5/2^+_2$ & 4.8 & -2.7  \\
$7/2^+_1$ & -23.2 & -22.5 & $7/2^+_2$ & -2.4 & -5.3  \\
$9/2^+_1$ & -17.2 & -19.5 & 
\end{tabular}
\caption{The electric quadrupole moment of the positive-parity states
 calculated by the AMD+GCM wave function ($Q$) and the rigid rotor
 approximation ($Q_s$).} \label{tab::Qmom}  
\end{ruledtabular}
\end{table}

Finally, we discuss observables and compare them with available
experimental data. The magnetic moments of the ground state ($1/2^+_1$)
and the first excited state ($3/2^+_1$) calculated without the spin
quenching factor are $-0.91\mu_N$ and $0.43\mu_N$, respectively. The
magnetic moment of the ground state shows the reasonable agreement with
the observed value $-0.89\mu_N$ \cite{Neyens} and the shell model
result\cite{Marechal}, while 
$3/2^+_1$ state has the opposite sign to the ground state.  
The $\log ft$ value of the $\beta$ decay from
$^{31}{\rm Na}$ and to $^{31}{\rm Al}$ are summarized in TABLE
\ref{tab::logft}. The ground state of $^{31}{\rm Na}$ and the low-lying states
of $^{31}{\rm Al}$ are also calculated by the AMD+GCM. The squared
overlaps between the AMD+GCM wave function and the AMD wave function 
are also listed to show the particle-hole nature of the low-lying states
$^{31}{\rm Mg}$ and $^{31}{\rm Al}$. It is found that
the ground state of $^{31}{\rm Na}$ ($3/2^+$) is the mixture of the neutron
$0p0h$, $2p2h$ and $4p4h$ configurations. The squared overlaps amount to
10\%, 70\% and 5\%, respectively. These configurations feed the $0p1h$,
$2p3h$ and $4p5h$ configurations of $^{31}{\rm Mg}$. Here, we must
recall that the ground state of $^{31}{\rm Mg}$ purely consist of the
$2p3h$ configuration. Therefore, it is populated only from the
$2p2h$ configuration of $^{31}{\rm Na}$ that is the dominant component
of the $^{31}{\rm Na}$. Indeed, the calculated $\log ft$ shows the
stronger $^{31}{\rm Na({\rm g.s.})}\rightarrow {}^{31}{\rm Mg}({\rm g.s.})$
transition than others. It is also notable that the excited states of 
$^{31}{\rm  Mg}$ dominated by the  $2p3h$ configuration have stronger
transition than those dominated by the $1p2h$ configuration. Though the
spin-parity assignment of the excited states are not fixed yet, the
tendency of the $\log ft$ is comparable with the experiment
\cite{Guillemaud,Klotz}. The $\beta$-decay from $^{31}{\rm Al}$ also
reflects the specific charactor of the $^{31}{\rm Mg}$. Since, the
ground state of $^{31}{\rm Mg}$ purely consist of the neutron $2p3h$
configuration, it feeds only the neutron $2p4h$ configuration of
$^{31}{\rm Al}$. The low-lying states of $^{31}{\rm Al}$ are dominated
by the neutron normal  configuration ($0p2h$) and the intruder
configuration ($2p4h$). The $1/2^+_2$, $3/2^+_2$, $3/2^+_3$, and
$1/2^+_3$ states dominated by the neutron $2p4h$ configuration have
smaller $\log ft$ than those dominated by other neutron
configurations. The strength of $\log ft$ and the excitation energy
shows good agreement with the experimental data \cite{Marechal}. The
spin-parity assignment of those $^{31}{\rm Al}$ states are not
experimental fixed yet, but the present result and the shell model
calculation \cite{Caurier2, Marechal} shows the reasonable  agreement. 
Table. \ref{tab::Qmom} summarizes the quadrupole moments of the
positive-parity states calculated by the AMD+GCM wave function and those
evaluated by the  rigid rotor approximation \cite{Bohr},  
\begin{eqnarray*}
 Q_s = \frac{3K^2-J(J+1)}{(J+1)(2J+3)}Q_0,
\end{eqnarray*}
where the intrinsic quadrupole moment $Q_0$ is calculated from the
intrinsic wave function $\Phi^\pi_{\rm int}(\beta)$ that is the dominant
component of each state. The K quantum number is taken as $1/2$ for the
states dominated by the $1p2h$ configuration and $3/2$ for those
dominated by the $3p4h$ configuration. Due to the strong
deformation, the yrast states that have the intruder
configuration show large collectivity. It is also notable that the
rigid rotor approximation qualitatively reproduces these states, even
though there are the configuration mixing in the AMD+GCM calculation.

To summarize, we have investigated the low-lying level structure of
$^{31}{\rm Mg}$. The observed property of the ground state including the
spin-parity is successfully reproduced. The coexistence of the spherical
normal configuration and the strongly deformed intruder configuration at
vary small excitation energy is predicted. The analysis of the single
particle orbital has shown that the neutron $0p1h$, $1p2h$, $2p3h$ and
$3p4h$ configurations appear on the energy surface. The angular momentum
projection leads to the coexistence of these configurations and the GCM
calculation has shown the mixing between them. Among the obtained
states, only the ground state consists of the pure neutron $2p3h$
configuration and its spin-parity $1/2^+$ originates in a neutron hole in
the Nilsson $[2,0,0,1/2]$ orbital. This specific charactor is reflected
to the $\beta$-decay strengths from the ground state of $^{31}{\rm Na}$
and to the low-lying states of $^{31}{\rm Al}$.

The author acknowledges helpful discussions with
Prof. Y. Kanada-En'yo. Most of the computational calculations are
carried out by the SX-5 super computer at Research Center for Nuclear
Physics, Osaka University (RCNP). This work is a part of the Research
Project for Study of Unstable Nuclei from Nuclear Cluster Aspects.

\bibliography{apssamp}

\end{document}